\documentclass{PoS}
\newcommand{\apj}{\emph{ApJ}}
\newcommand{\apjl}{\emph{ApJL}}
\newcommand{\apjs}{\emph{ApJS}}
\newcommand{\aap}{\emph{A\&A}}
\usepackage{graphicx}
\newcommand{\nat}{\emph{Nature}}

\title{The VERITAS Survey of the Cygnus Region of the Galaxy}

\ShortTitle{The Cygnus Region with VERITAS}

\author{\speaker{Alexis Popkow} for the VERITAS collaboration\thanks{veritas.sao.arizona.edu}\\
        University of California, Los Angeles\\
        E-mail: \email{apopkow@ucla.edu}}

\abstract{The Cygnus region is a very active region of our Galaxy with many sources of GeV and TeV gamma-ray emission, such as supernova remnants, pulsar wind nebulae, and massive star clusters. A detailed study of the Cygnus region at these energies can give insight into the processes of particle acceleration in astrophysical sources. VERITAS (Very Energetic Radiation Imaging Telescope Array System) is an array of four 12 meter diameter imaging atmospheric Cherenkov telescopes located at the Fred Lawrence Whipple Observatory (FLWO) in southern Arizona. From 2007 through 2012 VERITAS observed the Cygnus region for nearly 300 hours, from 67 to 82 degrees Galactic longitude and from -1 to 4 degrees in Galactic latitude. We have reanalyzed the VERITAS data with updated more sensitive analysis techniques and will be cross correlating that data with the results of an analysis of nearly six years of \textit{Fermi}-LAT data in the region. Using this cross correlation we can motivate continued observations in this active region of the Galaxy.}

\FullConference{The 34th International Cosmic Ray Conference,\\
		30 July- 6 August, 2015\\
		The Hague, The Netherlands}

\begin{document}

\section{Introduction}

Very-high-energy (VHE) gamma-ray sources are a fascinating probe into cosmic particle acceleration processes in the most extreme environments of the Milky Way Galaxy. There are a number of Galactic VHE source classes that have been detected including supernova remnants (SNRs), pulsar wind nebulae (PWNe), binary systems, and star forming regions, as well as yet unidentified sources of VHE emission.  SNRs and PWNe, in particular, are the likely accelerators of cosmic rays of Galactic origins. PWNe may also be the source of the positron excess detected by Pamela\cite{Pamela}. 
 
In recent years the H.E.S.S. Galactic Plane Survey (HGPS) has been a fruitful probe of the central area of the Galaxy \cite{HGPS}, detecting numerous new VHE sources, and probing diffuse Galactic VHE emission. The Cygnus region is a known region of VHE gamma ray emission, with emission detected by Milagro \cite{Milagro}, HEGRA \cite{hegra1} \cite{hegra2}, as well as VERITAS \cite{GammaCyg}\cite{OB1}\cite{TeV2032}. It also contains a number of potential gamma-ray emitters, identified by the EGRET \cite{EGRET} and  \textit{Fermi}-LAT GeV experiments \cite{3FGL} \cite{1FHL}. Motivated by the density of potential VHE gamma-ray emitters, VERITAS undertook survey observations of the region from 2007-2009. These observations, as well as follow-up observations have been analyzed to provide the deepest look at the Cygnus region at TeV energies to date. 

\section{VERITAS Survey and Follow-up Observations}

VERITAS is an imaging atmospheric Cherenkov telescope (IACT) array system located at the Fred Lawrence Whipple Observatory (FLWO) in southern Arizona ($31^{\circ} 40'$ N, $110^{\circ} 57'$ W, 1.3km a.s.l.). The array consists of four 12 m diameter telescopes with a field-of-view of $3.5^{\circ}$, which detect gamma rays at energies of 85 GeV to > 30 TeV at an energy resolution of 15-25\%. The pointing accuracy of the array is better than 50 arcsec and the angular resolution is < $0.1^{\circ}$ with a 68\% containment radius at 1 TeV. Four-telescope observations began in 2007, and there have been three major upgrades in the subsequent years. A telescope move in 2009 improved the sensitivity of the array \cite{T1}, and a trigger upgrade in winter 2012 and a camera upgrade in summer 2012 have improved the sensitivity further and lowered the energy threshold of the array.

\begin{figure}[h]
\centering
\includegraphics[trim=0cm 1cm 0cm 1cm, clip=true, width=1\textwidth]{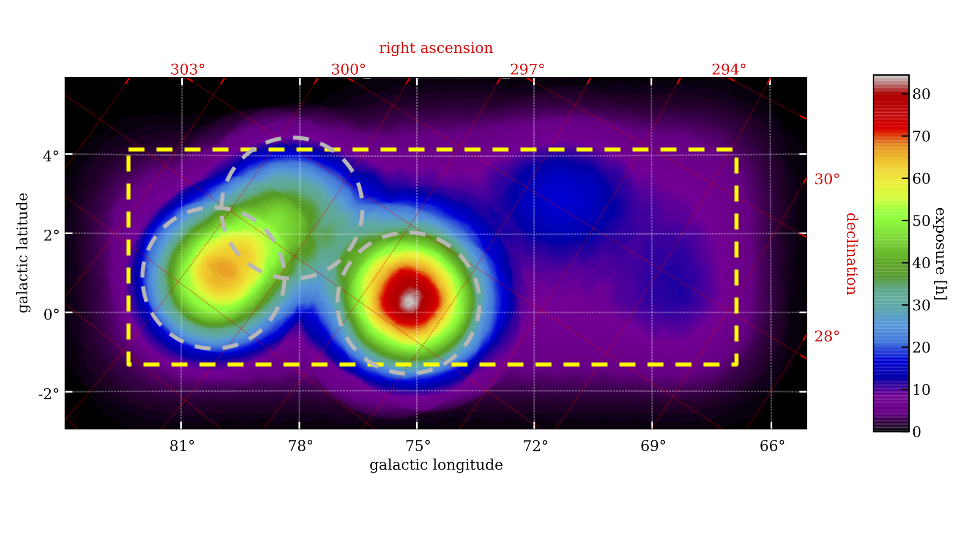}
\caption{\label{fig:Exposure} The radial acceptance corrected exposure map for the VERITAS observations, including follow-up observations. The scale on the color bar is in hours. The yellow box demarcates the core survey region, and the gray circles indicate the follow-up observation regions.}
\end{figure}

From 2007 through 2009, VERITAS undertook a survey of a $15^{\circ} \times 5^{\circ}$ area of the Cygnus region of the Galaxy, covering Galactic longitude (l) from $67^{\circ}$ to $82^{\circ}$ and Galactic latitude (b) from $-1^{\circ}$ to $4^{\circ}$. The survey consisted of more than 140 hours of observations, reaching an average point-source sensitivity of better than 4\% Crab Nebula flux at energies above 200 GeV \cite{Weinstein}. The base survey observations achieved a nearly uniform effective exposure of $\sim$6-7 hours at every location over the field of the survey. Observations were scheduled to keep the average zenith angle around $20^{\circ}$ as much as possible in order to try to avoid the higher energy threshold associated with large zenith angle observations. The full survey dataset also includes over 150 hours of follow-up and point source observations in the region (taken from 2008 through 2012). Follow-up data were taken using the standard VERITAS wobble observing strategy for point sources \cite{thesis}.

The previous blind search analysis from the survey plus follow-up observations yielded two source detections and suggested possible VHE $\gamma$-ray emission at several other locations. There were three regions observed in follow-up observations, shown in Figure \ref{fig:Exposure}. Four sources were detected: VER J2019+407 (G78.2+2.1, $\gamma$-Cygni) is likely a SNR shock front interacting with ambient material, and demonstrates  hadronic emission \cite{GammaCyg}, TeV J2032+413, the first unidentified TeV source, and likely PWN \cite{TeV2032}, and in the OB1 region (coincident with MGRO J2019+368) VER J2016+371 (a point source associated with CTB 87 and a likely PWN) and VER J2019+368 (a broad excess with some substructure suggesting the possibility of multiple sources) \cite{OB1}. We will be able to update and expand on these results with the improvements in the analysis over the last few years.

\section{Analysis}

\subsection{VERITAS Analysis}

The analysis of the Cygnus region data set begins in the same way as the standard VERITAS analysis with a calibration and cleaning of the shower images, followed by a principal component analysis of the shower images \cite{Daniel}. The trials factors of the analysis have been reduced by limiting the number of analyses taken with cuts. The analysis choices are motivated by the characteristics of know Galactic sources. The \textit{a priori} cuts include a choice of two spectral cuts: one for \textit{moderate} and one for \textit{hard} sources. The spectral cuts take into account the telescope upgrades, and are optimized for the spectral index and strength of the potential source, and for different energy thresholds. There is also an analysis with known sources are defined according to the published VERITAS analysis values for extension. The backgrounds are estimated using the ring background method applied at each skymap bin following the methods of \cite{Berge}.

In order to optimize the analysis sensitivity while limiting the number of trials over the large skymap we undertook an analytical calculation following \cite{Scott} to determine the optimum bin size for the analysis. Because Galactic sources are typically extended, we assumed a two-dimensional Gaussian with a $1\sigma$ (1 standard deviation) containment diameter of $0.5^{\circ}$ on a 10\% Crab source with 15 hours of exposure. The optimum bin size is linearly dependent on the width of the Gaussian and inversely proportional to the cube root of the intensity. The optimum bin size was found to be just a bit under $0.1^{\circ}$; thus, in this analysis the results are re-binned to $0.1^{\circ}$ prior to further analysis. 

Furthermore, in order to fully account for the trials in the analysis we are performing a Monte Carlo simulation \cite{Gross}. In each step of the simulation a map of randomly generated background gamma-rays is populated according to the dead-time and radial acceptance corrected exposure map. Then a ring background analysis is done on each bin of the map, and the highest significance bin and a random bin significance are recorded. In this analysis the random bin is picked to be in the core survey region, and not in the edges of the sky map. This is repeated about  $10^5$ times to get large the statistics to fit to high significances. The  trials factor of the VERITAS analysis will be taken into account in the final results.

\subsection{\textit{Fermi}-LAT Analysis}

\begin{figure}[h]
\centering
\includegraphics[trim=.5cm 0cm .5cm 1cm, clip=true, width=1\textwidth]{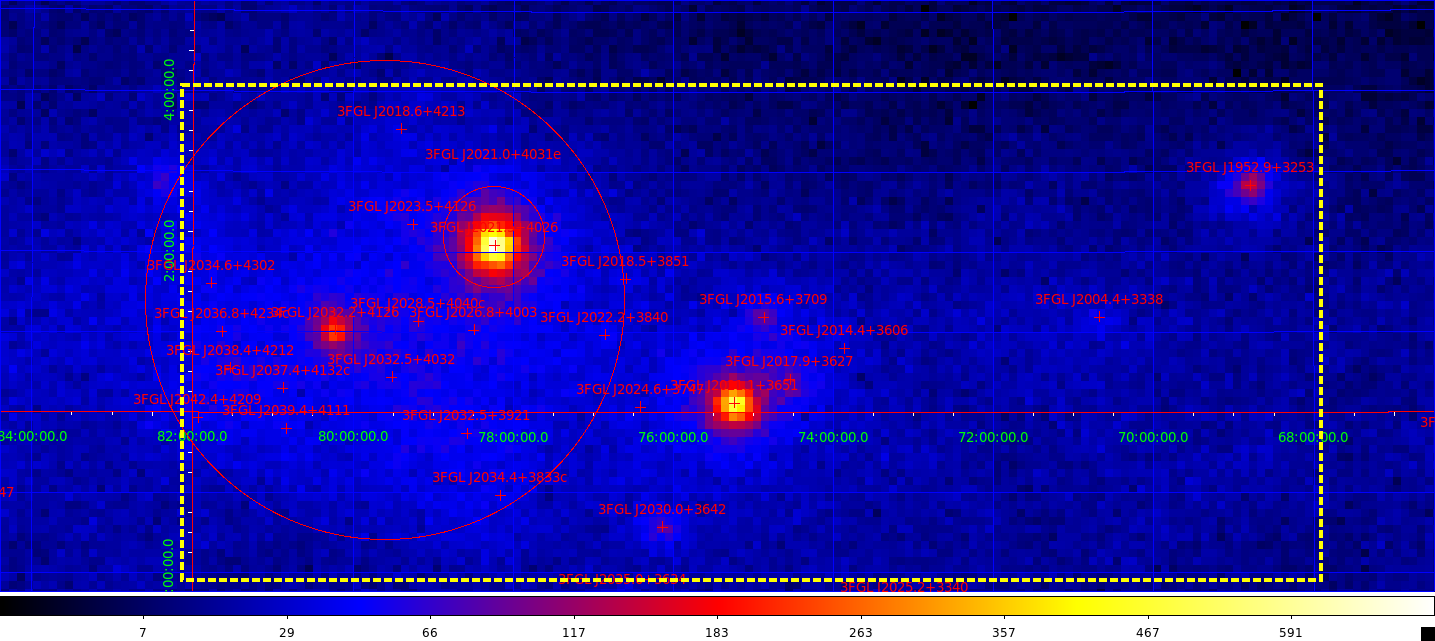}
\caption{\label{fig:FermiCyg}A \textit{Fermi} counts map of the Cygnus (>1 GeV) region using nearly six years of data, Fermi 3FGL sources are labeled. The square denotes the VERITAS sky survey region. The color bar scale is in square root of counts.}
\end{figure}

An analysis of six years of \textit{Fermi}-LAT data (Pass 7) in the Cygnus region in the energy range from 1 - 300 GeV has been undertaken using the publicly available LAT data tools \footnote{\textit{Fermi}-LAT tools v9r33p0. They are available at http://fermi.gsfc.nasa.gov/ssc/data/analysis/} and the binned analysis technique. Since this is a large region the analysis was initially broken up into three regions of $10^{\circ}$ radii centered at (l,b) of ($79.5^{\circ}$, $1.5^{\circ}$), ($74.5^{\circ}$, $1.5^{\circ}$), and ($69.5, 1.5^{\circ}$). Sources outside of $7^{\circ}$ and with low significances in the 3FGL catalog (<5$\sigma$) \cite{3FGL} were initially fixed. The results of these fits were then combined to a total fit of a $15^{\circ}$ region centered at the middle of the region. In each of these fits the extended sources in the region were held fixed and modeled using the templates provided by the \textit{Fermi}-LAT collaboration. A model and a counts map (Figure \ref{fig:FermiCyg}) were compared to check for any residuals, and the region is well fit by the model. The results from the fit are summarized in Table \ref{Fermi}.

There are 25 3FGL catalog sources in the VERITAS survey Cygnus region (labeled in Figure \ref{fig:FermiCyg}) including two that are low significance ($<5\sigma$) in the catalog, and two that are extended sources (Gamma Cygni, and the Cygnus Cocoon). The 21 remaining sources were fit using the binned analysis described above. Two of them are of potential extra-galactic origin: 3FGL J2015.6+3709 (associated with MG2 J201534+3710, an FSQR) and 3FGL J2018.5+3851 (associated with TXS 2016+386, an active galaxy of uncertain type). Of the sources with Galactic associations: 3FGL J2032.2+4126, is associated with the LAT pulsar J2032+4127, and is tentatively associated with the unidentified TeV source J2032+4130, which could have a PWN association, and 3FGL J2021.1+3651 and 3FGL J2017.9+3627 are associated with the region of the extended Milagro source MGRO J2019+37. Two of the remaining sources (3FGL J1952.9+3253 and 3FGL J2021.5+4026) are associated with pulsars, and the remaining 14 are unidentified. All of the sources that were fit in the analysis of the \textit{Fermi}-LAT data were studied for promising characteristics for potential TeV emission such as a strong GeV detection significance, a hard spectrum, and a high extrapolated flux. We will cross correlate the results and maps of this analysis with the VERITAS results.

Furthermore, there are nine sources within the VERITAS survey region in the First \textit{Fermi}-LAT Catalog of Sources above 10 GeV (designated 1FHL) \cite{1FHL}, which uses data from the first three years of the \textit{Fermi} mission. Included among these sources are two sources associated with active galactic nuclei, four pulsars with pulsation above 10 GeV (two of which are associated with the TeV sources MGRO J2019+37 and TeV J2032+4130), one GeV source associated with the SNR G078.2+02.1, one associated with the star forming region of the Cygnus Cocoon, and one unidentified source. Due to the longer observation window of our analysis and the longer time window of the 3FGL catalog, 3FGL catalog sources were used to populate the binned analysis inputs.

\begin{table}[]
\centering
\resizebox{\textwidth}{!}{%
\begin{tabular}{llllll}
\hline
{\bf 3FGL name} & {\bf Association(s)} & {\bf TS} & {\bf \begin{tabular}[c]{@{}l@{}}Flux (1 - 300 GeV) \\ ($cm^{-2} s^{-1}$)\end{tabular}} & {\bf Spectral shape} & {\bf \begin{tabular}[c]{@{}l@{}}Extrapolated Flux \\ (300 GeV - 1 TeV)\\ ($cm^{-2} s^{-1}$)\end{tabular}} \\ \hline
J1952.9+3253 & \begin{tabular}[c]{@{}l@{}}PSR J1952+3252\\  1FHL J1953.3+3251\end{tabular} & 5973 & 1.96E-08+/-4.17E-10 & broken power law & 1.19E-09 \\
J2004.4+3338 & 1FHL J2004.4+3339 & 246 & 3.48E-09+/-3.04E-10 & power law & 6.97E-13 \\
J2015.6+3709 & \begin{tabular}[c]{@{}l@{}}MG2 J201534+3710 \\ 1FHL J2015.8+3710\end{tabular} & 908 & 9.06E-09+/-4.30E-10 & log parabola & 1.11E-13 \\
J2017.9+3627 & MGRO J2019+37 & 667 & 9.84E-09+/-5.16E-10 & log parabola & 1.72E-15 \\
J2018.5+3851 & \begin{tabular}[c]{@{}l@{}}TXS 2016+386\\  1FHL J2018.3+3851\end{tabular} & 76 & 1.88E-09+/-3.09E-10 & power law & 7.39E-13 \\
J2018.6+4213 &  & 34 & 1.41E-09+/-2.90E-10 & power law & 1.27E-12 \\
J2021.1+3651 & \begin{tabular}[c]{@{}l@{}}PSR J2021+3651\\  MGRO J2019+37\\  1FHL J2021.0+3651\end{tabular} & 23561 & 6.84E-08+/-8.04E-10 & broken power law & 4.15E-10 \\
J2021.5+4026 & LAT PSR J2021+4026 & 46339 & 1.13E-07+/-9.45E-10 & broken power law & 2.79E-10 \\
J2022.2+3840 &  & 98 & 3.35E-09+/-3.90E-10 & log parabola & 4.50E-20 \\
J2023.5+4126 &  & 84 & 3.29E-09+/-4.25E-10 & log parabola & 3.52E-14 \\
J2024.6+3747 &  & 28 & 1.37E-09+/-4.78E-10 & log parabola & 3.69E-15 \\
J2026.8+4003 &  & 68 & 3.19E-09+/-4.46E-10 & log parabola & 1.13E-14 \\
J2032.2+4126 & \begin{tabular}[c]{@{}l@{}}LAT PSR J2032+4127 \\ TeV J2032+4130\\  1FHL J2032.1+4125\end{tabular} & 4717 & 2.45E-08+/-5.74E-10 & broken power law & 1.23E-09 \\
J2032.5+3921 &  & 132 & 4.38E-09+/-4.32E-10 & log parabola & 6.70E-16 \\
J2032.5+4032 &  & 116 & 4.20E-09+/-6.33E-10 & log parabola & 7.13E-22 \\
J2034.6+4302 &  & 69 & 2.65E-09+/-5.01E-10 & log parabola & 1.60E-18 \\
J2036.8+4234c &  & 75 & 2.52E-09+/-4.38E-10 & power law & 4.53E-14 \\
J2037.4+4132c &  & 62 & 2.42E-09+/-4.24E-10 & power law & 1.19E-13 \\
J2038.4+4212 &  & 129 & 3.64E-09+/-4.68E-10 & log parabola & 3.20E-14 \\
J2039.4+4111 &  & 75 & 2.71E-09+/-3.98E-10 & log parabola & 1.97E-14 \\
J2042.4+4209 &  & 76 & 2.68E-09+/-3.82E-10 & log parabola & 4.70E-14
\end{tabular}%
}
\caption{The table summarizes the results of the fit of six years of Fermi-LAT data in the Cygnus region. The 21 sources used for the binned analysis are listed together with its associated source(s) if applicable and a test statistic (TS) value, the flux from 1 - 300 GeV, and an extrapolated flux at 300 GeV - 1 TeV is calculated using the spectral shape noted in the table.}
\label{Fermi}
\end{table}

%\section{Results}

\section{Conclusions and Outlook}

The VERITAS survey of the Cygnus region serves to enhance the knowledge about the populations of VHE gamma-ray emitters, and potential VHE gamma-ray emitters in our Galaxy, using over 300 hours of observations in the region from $67^{\circ}$ to $82^{\circ}$ Galactic longitude and from $-1^{\circ}$ to $4^{\circ}$ Galactic latitude. This region has been determined to be active in VHE gamma rays with four sources detected by VERITAS and is furthermore a region for active inquiry at GeV energies with a large number of unidentified sources. The new VERITAS analysis in the region has increased sensitivity, and can help untangle the origins of gamma-ray emission in the region.

In the future, there will likely be further VERITAS observations in the region as a result of this analysis. The cross correlation of the \textit{Fermi} and VERITAS sky maps will provide strong motivation for low significance source candidates, and will allow for the possible identification or association of unidentified GeV sources in the region. Furthermore HAWC will have good sensitivity in the the region and VERITAS will work to follow-up on potential new sources detected with HAWC. Together this will enable a greater understanding of the region at the highest energies.

\acknowledgments{\noindent This research is supported by grants from the U.S. Department of Energy Office of Science, the U.S. National Science Foundation and the Smithsonian Institution, and by NSERC in Canada. We acknowledge the excellent work of the technical support staff at the Fred Lawrence Whipple Observatory and at the collaborating institutions in the construction and operation of the instrument. The VERITAS Collaboration is grateful to Trevor Weekes for his seminal contributions and leadership in the field of VHE gamma-ray astrophysics, which made this study possible.}


\begin{thebibliography}{99}

\bibitem{Pamela} O. Adriani, G.~C. Barbarino, G.~A. Bazilevskaya et al., \emph{An anomalous positron abundance in cosmic rays with energies $1.5-100$ GeV}, \nat, 458, 607 (2009).

\bibitem{HGPS} S. Carrigan, F. Brun, R.~C.~G. Chaves, et al., \emph{The H.E.S.S. Galactic Plane Survey - maps, source catalog and source population}, in the proceedings of \emph{The 33rd International Cosmic Ray Conference} (2013) [{\tt astro-ph/1307.4690}].

\bibitem{Milagro} A.~A. Abdo, B. Allen, D. Berley, et al., \emph{TeV Gamma-Ray Sources from a Survey of the Galactic Plane with Milagro}, \apjl, 664, L91, (2007) [{\tt astro-ph/0705.0707}].

\bibitem{hegra1} F. Aharonian, A. Akhperjanian, M. Beilicke, et al., \emph{An unidentified TeV source in the vicinity of Cygnus OB2}, \aap, 393, L37 (2002) [{\tt astro-ph/0207528}].

\bibitem{hegra2} F. Aharonian, A. Akhperjanian, M. Beilicke, et al., \emph{The unidentified TeV source (TeV J2032+4130) and surrounding field: Final HEGRA IACT-System results}, \aap, 431, 197 (2005).

\bibitem{GammaCyg} E. Aliu, S. Archambault, T. Arlen, et al., \emph{Discovery of TeV Gamma-Ray Emission Toward Supernova Remnant SNR G78.2+2.1} \apj, 770, 93 (2013) [{\tt astro-ph/1305.6508}].

\bibitem{OB1} E. Aliu, T. Aune, B. Behera, et al., \emph{Spatially Resolving the Very High Energy Emission From MGRO J2019+37 with VERITAS}, \apj, 788, 78 (2014) [{\tt astro-ph/1404.1841}].

\bibitem{TeV2032} E. Aliu, T. Aune, B. Behera, et al., \emph{Observations of the Unidentified Gamma-ray Source TeV J2032+4130 by VERITAS}, \apj, 783, 16 (2014) [{\tt astro-ph/1401.2828}].

\bibitem{EGRET} R.~C. Hartman, D.~L.
Bertsch, S.~D. Bloom, et al., \emph{The Third EGRET Catalog of High-Energy Gamma-Ray Sources} , \apjs, 123, 79 (1999).

\bibitem{3FGL} The Fermi-LAT Collaboration, \emph{Fermi Large Area Telescope Third Source Catalog} (2015) [{\tt astro-ph/1501.02003}].

\bibitem{1FHL} M. Ackermann, M. Ajello, A. Allafort, et al., \emph{The First Fermi-LAT Catalog of Sources Above 10 GeV}, \apjs, 209, 34 (2013) [{\tt astro-ph/1306.6772}].

\bibitem{T1} A.~N. Otte for the VERITAS collaboration, \emph{Upgrade of the VERITAS Cherenkov Telescope Array} (2009) [{\tt astro-ph/0907.4826}].

\bibitem{Weinstein} A. Weinstein for the VERITAS Collaboration, \emph{The VERITAS Survey of the Cygnus Region of the Galactic Plane} (2009) [{\tt astro-ph/0912.4492}].

\bibitem{thesis} A. W. Smith, \emph{Multiwavelength Observations of the TeV Binary LS I +61 303}, \emph{The University of Leeds}, Leeds, 2007.

\bibitem{Daniel} M.~K. Daniel, \emph{The VERITAS Standard Data Analysis}. \emph{The 30th International Cosmic Ray Conference} (2008) 3, 1325[{\tt astro-ph/0709.4006}].

\bibitem{Berge} D. Berge, S. Funk, \& J. Hinton, \emph{Background modelling in very-high-energy gamma-ray astronomy}, \aap, 466, 1219 (2007) [{\tt astro-ph/0610959}].

\bibitem{Scott} D. W. Scott, \emph{On Optimal and Data Based Histograms},\emph{Biometrika}, Vol. 66, No. 3 (Dec., 1979), pp. 605-610.

\bibitem{Gross} E. Gross, O. Vitells, \emph{Trial factors for the look elsewhere effect in high energy physics}, \emph{The European Physical Journal C}, 70, 525 (2010) [{\tt physics/1005.1891}].

\end{thebibliography}
\end{document}